\documentclass[english,format=sigconf]{extarticle}
\usepackage[T1]{fontenc}
\usepackage[latin9]{inputenc}
\usepackage{geometry}
\PassOptionsToPackage{ruled}{algorithm2e}
\usepackage{color}
\usepackage{babel}
\usepackage{array}
\usepackage{dvipost}
\usepackage{algorithm2e}
\usepackage{amsbsy}
\usepackage{graphicx}
\usepackage[numbers]{natbib}
\usepackage[unicode=true,pdfusetitle,
 bookmarks=true,bookmarksnumbered=false,bookmarksopen=false,
 breaklinks=false,pdfborder={0 0 1},backref=false,colorlinks=true]
 {hyperref}
\usepackage{breakurl}

\makeatletter

\newcommand*\LyXZeroWidthSpace{\hspace{0pt}}
\providecommand{\tabularnewline}{\\}
\dvipostlayout
\dvipost{osstart color push Red}
\dvipost{osend color pop}
\dvipost{cbstart color push Blue}
\dvipost{cbend color pop}

\SetAlFnt{\small}
\SetAlCapFnt{\small}
\SetAlCapNameFnt{\small}
\SetAlCapHSkip{0pt}
\IncMargin{-\parindent}

\usepackage{fancyhdr}
\LinesNumbered

\@ifundefined{showcaptionsetup}{}{%
 \PassOptionsToPackage{caption=false}{subfig}}
\usepackage{subfig}
\makeatother

\begin{document}

\title{Query-Efficient Black-Box Attack Against Sequence-Based Malware Classifiers}

\author{Ishai~Rosenberg, Asaf~Shabtai, Yuval~Elovici and Lior~Rokach\\
Ben-Gurion University of The Negev}
\maketitle
\begin{abstract}
In this paper, we present a generic, query-efficient black-box attack
against API call-based machine learning malware classifiers. We generate
adversarial examples by modifying the malware's API call sequences
and non-sequential features (printable strings), and these adversarial
examples will be misclassified by the target malware classifier without
affecting the malware's functionality. \\
In contrast to previous studies, our attack minimizes the number of
malware classifier queries required. In addition, in our attack, the
attacker must only know the class predicted by the malware classifier;
attacker knowledge of the malware classifier's confidence score is
optional. We evaluate the attack effectiveness when attacks are performed
against a variety of malware classifier architectures, including recurrent
neural network (RNN) variants, deep neural networks, support vector
machines, and gradient boosted decision trees. Our attack success
rate is around 98\% when the classifier's confidence score is known
and 64\% when just the classifier's predicted class is known. \\
We implement four state-of-the-art query-efficient attacks and show
that our attack requires fewer queries and less knowledge about the
attacked model's architecture than other existing query-efficient
attacks, making it practical for attacking cloud-based malware classifiers
at a minimal cost.
\end{abstract}

\section{\label{sec:Introduction}Introduction}

Next generation anti-malware products use machine learning and deep
learning models instead of signatures and heuristics, in order to
detect previously unseen malware. Windows-based API call sequence-based
classifiers, such as \citep{SentinelOne} or \citep{MicrosoftATP},
provide state-of-the-art detection performance \citep{DBLP:journals/corr/abs-1806-10741}.
However, those classifiers are vulnerable to adversarial example attacks.
Adversarial examples are samples that are perturbed (modified), so
they will be incorrectly classified by the target classifier. 

In this paper, we demonstrate a novel \emph{query-efficient} black-box
attack against many types of malware classifiers, including RNN variants.
We implement our attack on malware classifiers that are used to classify
a process as malicious or benign. This classification is based on
both API call sequences and additional discrete features (e.g., printable
strings). We consider the challenging case of machine learning as
a service (MLaaS). Examples of such services are Amazon ML \citep{AmazonMachineLearning}
and GCP \citep{GoogleCloudPrediction}. 

In this case, the attacker continuously queries the target malware
classifier (e.g., \citep{JoeSandboxML}) and modifies the API call
sequences the malware generates until the sequence is classified as
benign. The attacker pays for every query of the target malware classifier
and therefore aims to minimize the number of queries made to such
cloud services when performing an attack. Another reason for minimizing
the number of queries is that many queries from the same computer
might arouse suspicion of an adversarial attack attempt, causing the
cloud service to stop responding to those queries, e.g., using stateful
defenses that keep track of queries to the system \citep{chen2019stateful}.
While the attacker may use a botnet to issue many queries, this approach
would increase the attack's cost dramatically. 

We develop an \emph{end-to-end attack} \citep{rosenberg2020adversarial,RosenbergIJCNN20}
(a.k.a, problem-space attack \citep{DBLP:conf/sp/PierazziPCC20})
by recrafting the malware behavior so it can evade detection by such
machine learning malware classifiers while minimizing the amount of
queries to the malware classifier (meaning that the attack is \emph{query-efficient}).

The main focus of most existing research (e.g., \citep{DBLP:conf/icml/IlyasEAL18,DBLP:conf/icml/UesatoOKO18})
is on the query-efficient generation of adversarial examples for images.
This is different from our work, which is focused on generating adversarial
API sequences, in two respects:
\begin{enumerate}
\item In the case of adversarial API sequences, one must verify that the
original functionality of the malware remains intact. Thus, one cannot
simply generate an adversarial feature vector but must generate an
executable file containing the corresponding working malware behavior.
\item API sequences consist of discrete symbols of variable lengths, while
images are represented as matrices with fixed dimensions, and the
values of the matrices are continuous.
\end{enumerate}
In this paper, we modify the malware's behavior by modifying the API
call sequences it generates. We also modify static, non-sequential
features: printable strings inside the process (Appendix \ref{sec:Appendix-D:-Handling}).
We present eight novel attacks that are different in terms of three
parameters:
\begin{enumerate}
\item Attacker knowledge - The attacker might have knowledge about the target
classifier's confidence score (\emph{score-based attack}) or only
about the predicted class (\emph{decision-based attack}). This knowledge
can affect the way we modify both the positions and the types of modified
API calls.
\item Modified API call types (values) - The attacker can either modify
API calls randomly (random perturbation) or take them from a generative
adversarial network (GAN) generating benign samples (benign perturbation).
\item Method to select the number of modified API calls - The attacker can
either modify API calls until the sample successfully evades detection
with minimal perturbation (linear iteration attack) or start with
a maximum number of modifications and minimize them as long as evasion
is maintained (logarithmic backtracking attack).
\end{enumerate}
Each of the eight attacks we present is a combination of these three
parameters, each of which has two different variants (score-based
or decision-based; benign perturbation or random perturbation; logarithmic
backtracking or linear iteration). As a benchmark, we adapted four
state-of-the-art query-efficient attacks. We showed that our attacks
are equal or outperform them, obtaining a higher success rate (98\%
using the classifier's confidence score and 88\% without it) for a
fixed number of queries.

The contributions of our paper are as follows:
\begin{enumerate}
\item This is the first \emph{query-efficient, end-to-end, black-box} adversarial
attack for \emph{both sequential and non-sequential} input. 
\item We provide two variants for our attack, to fit the knowledge available
for the attacker (confidence score or label only), thus fitting practical
cyber security domain use cases.
\item This is the first usage of GAN (previously used for other purposes)
to generate benign API call trace samples to produce a query-efficient
attack.
\item This is the first usage of self adaptive evolutionary algorithm (previously
used for other purposes) to implement a query-efficient score-based
attack.
\end{enumerate}

\section{\label{sec:Background-and-Related}Background and Related Work}

The search for adversarial examples, such as those used in our attack,
can be formalized as a minimization problem: 

\begin{equation}
\arg_{\boldsymbol{\mathbf{r}}}\min f(\boldsymbol{\mathbf{x}}+\boldsymbol{\mathbf{r}})\neq f(\boldsymbol{\mathbf{x}})\:s.t.\:\boldsymbol{\mathbf{x}}+\boldsymbol{\mathbf{r}}\in\boldsymbol{\mathbf{D}}
\end{equation}

The input \textbf{$\boldsymbol{\mathbf{x}}$}, correctly classified
by the classifier $f$, is perturbed with \textbf{$\boldsymbol{\mathbf{r}}$},
such that the resulting adversarial example \textbf{$\boldsymbol{\mathbf{x}}+\boldsymbol{\mathbf{r}}$}
remains in the input domain \textbf{$\boldsymbol{\mathbf{D}}$} but
is assigned a different label than \textbf{$\mathbf{\boldsymbol{x}}$}.

There are three types of adversarial example generation methods:

In \emph{gradient-based attacks}, adversarial perturbations are generated
in the direction of the gradient, that is, in the direction with the
maximum effect on the classifier's output (e.g., fast gradient sign
method \citep{Goodfellow14}). Hu and Tan \citep{DBLP:journals/corr/HuT17a}
used RNN GAN to generate invalid API calls and insert them into the
original API sequences. Gumbel-Softmax, a one-hot continuous distribution
estimator, was used to deliver gradient information between the generative
RNN and substitute RNN. A white-box gradient-based attack against
RNNs demonstrated against long short-term memory (LSTM) architecture
for sentiment classification of a movie review dataset was shown in
\citep{Papernot2016a}. A black-box variant, which facilitates the
use of a substitute model, was presented in \citep{DBLP:conf/raid/RosenbergSRE18}.
The attack in this paper is different in a few ways: 
\begin{enumerate}
\item As shown in Section \ref{subsec:Attack-Performance}, the attack described
in \citep{DBLP:conf/raid/RosenbergSRE18} requires more target classifier
queries and greater computing power to generate a substitute model.
\item We use a different adversarial example generation algorithm, which
uses a stochastic approach rather than a gradient-based approach,
making it harder to defend against (as mentioned in Section \ref{subsec:Defenses-and-Mitigation}).
\item Our decision-based attack is generic and doesn't require a per malware
pre-deployment phase to generate the adversarial sequence (either
using a GAN, as in \citep{DBLP:journals/corr/HuT17a}, or a substitute
model, as in \citep{DBLP:conf/raid/RosenbergSRE18}). Moreover, generation
takes place at run time, making it even more generic and easier to
deploy.
\end{enumerate}
\emph{Score-based attacks} are based on knowledge of the target classifier's
confidence score. Previous research used a genetic algorithm (GA),
where the fitness of the genetic variants is defined in terms of the
target classifier\textquoteright s confidence score, to generate adversarial
examples that bypass PDF malware and image recognition classifiers,
respectively \citep{DBLP:conf/ndss/XuQE16,DBLP:journals/corr/abs-1805-11090}.
Those attacks used a computationally expensive GA compared to our
approach and were only evaluated when performed against support vector
machine (SVM), random forest, and CNN classifiers using static features
only, and was not evaluated against recurrent neural network variants
using both static and dynamic analysis features, as we do. In \citep{DBLP:conf/icml/UesatoOKO18},
the simultaneous perturbation stochastic approximation (SPSA) method
was used, since its gradient approximation requires only two queries
(for the target classifier's score) per iteration, regardless of the
dimension of the optimization problem. Alzantot et al. \citep{DBLP:conf/emnlp/AlzantotSEHSC18}
presented a sequence-based attack algorithm that exploits population-based
gradient-free optimization via GA; in this study, the attack was performed
against a natural language processing (NLP) sentiment analysis classifier.
While this attack can be used against RNNs, it requires more queries
than our attack (see Table \ref{tab:Attack-Comparison}).

\emph{Decision-based attacks} only use the label predicted by the
target classifier. Ilyas et al. \citep{DBLP:conf/icml/IlyasEAL18}
used natural evolutionary strategies (NES) optimization to enable
query-efficient gradient estimation, which leads to the generation
of misclassified images as seen in gradient-based attacks. Dang et
al. \citep{DBLP:conf/ccs/DangHC17} used the rate of feature modifications
from known malicious and benign samples as the score and used a hill
climbing approach to minimize this score, evading SVM and random forest
PDF malware classifiers based on static features in an efficient manner.
Our approach, on the other hand, is more generic and can handle RNN
classifiers and multiple feature types. The performance differences
between our approaches and \citep{DBLP:conf/ccs/DangHC17} are presented
in Section \ref{subsec:Comparison-to-Previous}.

All of the currently published score and decision-based attacks differ
from our proposed attack in that: 
\begin{enumerate}
\item They \emph{only} deal with CNNs, random forest, and SVM classifiers,
using non-sequential input, as opposed to \emph{all} state-of-the-art
classifiers (including RNN variants), using discrete or sequence input,
as in our attack. 
\item They deal primarily with images and rarely fit the attack requirements
of the cyber security domain: while changing a pixel's color doesn't
``break'' the image, modifying an API call might harm the malware
functionality. In addition, small perturbations, such as those suggested
in \citep{DBLP:conf/icml/IlyasEAL18,DBLP:conf/icml/UesatoOKO18},
are not applicable for discrete API calls: you can't change \emph{WriteFile()}
to \emph{WriteFile()}+0.001 in order to estimate the gradient to perturb
the adversarial example in the right direction; you need to modify
it to an entirely different API. This is reflected in Table \ref{tab:Attack-Comparison}.
\item They did not present an end-to-end framework to implement the attack
in the cyber security domain. Thus, the attack might be used for generating
adversarial malware feature vectors but not for generating a working
adversarial malware sample.
\end{enumerate}
The differences between those attacks and our attacks are summarized
in Table \ref{tab:Comparison-to-Previous}. 

\begin{table*}
\caption{\label{tab:Comparison-to-Previous}Comparison to Previous Work}

\centering{}%
\begin{tabular}{|>{\centering}p{0.2\textwidth}||>{\centering}p{0.15\textwidth}|>{\centering}p{0.3\textwidth}|>{\centering}p{0.1\textwidth}|>{\centering}p{0.12\textwidth}|}
\hline 
Attack Type & Domain  & Input Type & Query-Efficient? & Score/Decision-Based?\tabularnewline
\hline 
\hline 
Rosenberg et al. \citep{DBLP:conf/raid/RosenbergSRE18} (Gradient-Based
Attack) & Cyber Security & Sequence, Non-sequential, Mixed & No & Decision\tabularnewline
\hline 
\hline 
Uesato et al. \citep{DBLP:conf/icml/UesatoOKO18} & Image Recognition & Non-sequential & Yes & Score\tabularnewline
\hline 
\hline 
Ilyas et al. \citep{DBLP:conf/icml/IlyasEAL18} & Image Recognition & Non-sequential & Yes & Score\tabularnewline
\hline 
\hline 
Alzantot et al. \citep{DBLP:journals/corr/abs-1805-11090} & NLP & Sequence & Yes & Score\tabularnewline
\hline 
\hline 
Our Score-Based Attack & Cyber Security & Sequence, Non-sequential, Mixed & Yes & Score\tabularnewline
\hline 
\hline 
Our Decision-Based Attack & Cyber Security & Sequence, Non-sequential, Mixed & Yes & Decision\tabularnewline
\hline 
\end{tabular}
\end{table*}

\section{Methodology}

\subsection{\label{subsec:Target-API-Call}Attacking API Call-Based Malware Classifiers}

An overview of the malware classification process is shown in Figure
\ref{fig:Overview-of-the} (in the appendix). 

Assume a malware classifier whose input is a sequence of API calls
made by the inspected process. API call sequences can be millions
of API calls long, making it impossible to train such a classifier
on the entire sequence at once, due to training time and GPU memory
constraints. Thus, the target classifier uses a non-overlapping sliding
window approach \citep{DBLP:conf/raid/RosenbergSRE18}: Each API call
sequence is divided into windows of $k$ API calls. Detection is performed
on each window in turn, and if any window is classified as malicious,
the entire sequence is considered malicious. Thus, even cases such
as malicious payloads injected into goodware (e.g., using Metasploit),
where only a small subset of the sequence is malicious, would be detected. 

We use one-hot encoding for each API call type in order to cope with
the limitations of scikit-learn's implementation of decision trees
and random forests, as mentioned in \citep{SciKitLearnDecisionTreeCategorialVar}.
The output of each classifier is either the predicted class (whether
the API call trace is malicious or benign) or the confidence score
of the prediction (a value between 0 for a benign process to 1.0 for
a malicious process). Appendix \ref{sec:Appendix-B:-Tested} contains
a description of the classifiers used in our study and their hyperparameters.

In order to attack such a malware classifier, we want to add API calls
without changing the malware functionality. Removing API calls without
modifying the malware functionality requires complex analysis of the
malware code and thus cannot be scaled. Therefore, we use a \emph{no-op
mimicry attack} \citep{Wagner2002}, that is, we add API calls with
no effect or an irrelevant effect on the malware functionality. We
use this method regardless of the perturbation used to generate the
API call type, either random or benign perturbations (shown below).
Almost every API call can become a no-op if provided with the right
arguments, e.g., opening a non-existent file. 

However, analyzing arguments would make our attack easier to detect,
e.g., by considering only successful API calls and ignoring failed
API calls or by looking for irregularities in the arguments of the
API calls (e.g., invalid file handles, etc.). In order to address
this issue, we use valid (non-null) arguments with a no-op effect,
such as writing into a temporary file handle (instead of an invalid
file handle) or reading zero bytes from a valid file. This makes detecting
the no-op API calls much harder, since the API call runs correctly,
with a return value indicative of success. It is extremely challenging
for the malware classifier to differentiate between malware that is
trying to read a non-existent registry key as an added adversarial
no-op and a benign application functionality, e.g., trying to find
a registry key containing information from previous runs and creating
it if it doesn't exist (for instance, during the first run of the
application). This makes our problem-space attack robust to preprocessing
\citep{DBLP:conf/sp/PierazziPCC20}.

To conclude, we believe that any detector will suffer from a very
high false positive rate and thus will not be a practical solution
for detecting our attacks. We don't add API types which cannot be
no-opped and thus always affect the malware functionality. In this
study, we focus on the 314 API call types monitored by Cuckoo Sandbox.
Of those, only two API call types are not added because they can\textquoteright t
be no-opped: \emph{ExitWindowsEx}() and \emph{NtTerminateProcess}().
Every other API call have a no-op variant, generated by the abovementioned
method.

\subsection{\label{subsec:Black-Box-API-Call}The Proposed Black-Box Attack}

\begin{figure*}[tp]
\begin{centering}
\textsf{\includegraphics[scale=0.5]{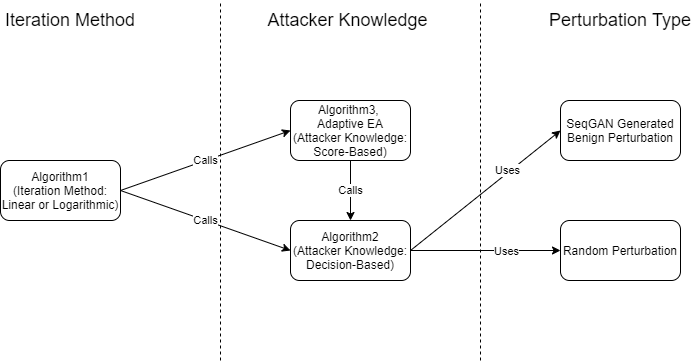}}
\par\end{centering}
\caption{\label{fig:The-Proposed-Attack's}The Proposed Attack's Flow}
\end{figure*}

Our proposed attack's flow is detailed in Figure \ref{fig:The-Proposed-Attack's}.
As mentioned before, our proposed attacks can be characterized by
the knowledge the attacker has, the method for selecting which new
API call to add (termed \emph{perturbation type}), and the method
to select the number of added API calls (termed \emph{iteration method}).
As can be seen in Figure \ref{fig:The-Proposed-Attack's}, these characteristics
affect the attack's flow. These characteristics are described in the
subsections below.

\subsubsection{\label{subsec:Full-Sequence-Linear}Iteration Method}

In this subsection, we describe the method to select the number of
added API calls by the attack. In addition to the regular \emph{linear
iteration} method, we propose an efficient \emph{logarithmic backtracking}
transformation as a method for determining the number of API calls
to add. This method starts with a large ratio of perturbation (that
is, a larger number of API calls are added to the original sequence
to fool the classifier) which rapidly decreases as long as the sequence
remains misclassified.

In order to handle the entire API call sequence, we use Algorithm
\ref{alg:Logarithmic-Backtracking-Attack}. In this algorithm, the
attacker splits the malicious API call sequence \textbf{$\boldsymbol{x^{m}}$}
into windows of $n$ API calls ($\boldsymbol{w_{j}^{m}}$), similar
to the division made by the malware classifier, and modifies each
window in turn using Algorithm \ref{alg:Full-Window-Sequence} (described
below; line 8). The API calls ``pushed out'' from $\boldsymbol{w_{j}^{m}}$
will become the beginning of $\boldsymbol{w_{j+1}^{m}}$, so no API
is ignored. The adversarial window size $n$ might be different from
the malware classifier's window size $k$, which is not known to the
attacker. As shown in Section \ref{subsec:Attack-Performance}, this
has a negligible effect on the attack performance. In the case of
a benign perturbation, the benign API call sequence \textbf{$\boldsymbol{x^{b}}$}
is similarly split into windows of $n$ API calls ($\boldsymbol{w_{j}^{b}}$).

The adversarial API call sequence length $l$ might be larger than
$n$, the length of the sliding window API call sequence that is used
by the adversary. Therefore, the attack is performed sequentially
on $\left\lceil \frac{l}{n}\right\rceil $ windows of $n$ API calls
(e.g., for $l=1000$ and $n$=100 the malware classifier would run
on ten windows, with API call indices: 1..100, 101..200, ..., 900..1000).
Note that the knowledge of $k$ (the window size of the malware classifier)
is not required, as shown in Section \ref{subsec:Attack-Performance}. 

$D$ is the vocabulary of available features. In this case, these
features are all of the API call types recorded by the malware classifier,
e.g., \emph{CreateFileW}(). Note that $D$ is not necessarily known
to the attacker. The attacker knows $D^{'}$ , which might be a subset
or superset of $D$. This knowledge of $D^{'}$ is a commonly accepted
assumption about the attacker's knowledge \citep{DBLP:conf/ccs/HuangJNRT11}.
In fact, it is enough for the attacker to know the feature type used
by the target classifier (API call types in this study), which is
public information that is usually published by classifier implementers.
With that knowledge, the attacker can cover \emph{all} API call types
(several thousands) to generate $D^{'}$ , which is a superset of
$D$. 

In our research, we observed that API call types in $D'-D$ are not
monitored by the classifier, and thus adding them does not assist
in creating adversarial examples; those API calls just add API call
overhead to the modified sequence and serve as wasted queries. API
call types in $D-D'$, unknown to the attacker, are not generated
by the attack and therefore decreasing the adversarial feature space
and thus decreasing the possibilities for generating modified sequences
that can evade detection. Thus, when $D^{'}$ is a superset of $D$,
the attack has higher overhead but remains as effective. An attacker
might also reverse engineer the features from the malware classifier
program, as has already done in real-world adversarial example generation
for malware classifier scenarios \citep{CylanceAdvAttack}.

In order to decrease the number of malware classifier queries (the
number of calls to $f(.)$), we can use \emph{logarithmic backtracking}.
This iteration method is similar to binary search. In this case, we
only query the malware classifier in Algorithm \ref{alg:Logarithmic-Backtracking-Attack}
after modifying $M_{w}$ API calls in Algorithm \ref{alg:Random-Window-Sequence}
(instead of querying the model after modifying a single API call in
a linear iteration), which should be a sufficiently large perturbation
to evade the malware classifier. Then, we start reducing the number
of modified API calls by half before querying the malware classifier
(lines 13-14) until it detects the sample again. Finally, we keep
restoring half of the API calls we previously removed before querying
(line 22), until we achieve a perturbation that fools the malware
classifier with a minimal number of additional API calls and malware
classifier queries.

\begin{algorithm*}
\caption{\label{alg:Logarithmic-Backtracking-Attack}Full Sequence Attack}

\noindent \begin{raggedright}
\textbf{Input}: $f$ - black-box model, 
\par\end{raggedright}
\noindent \begin{raggedright}
\textbf{$\boldsymbol{x^{m}}$} - malicious sequence to perturb, \textbf{$\boldsymbol{x^{b}}$}
- benign sequence to mimic, 
\par\end{raggedright}
\noindent \begin{raggedright}
$n$ - size of adversarial sliding window, $D^{'}$ - adversarial
vocabulary, 
\par\end{raggedright}
\noindent \begin{raggedright}
$M_{w}$ - maximum API modifications per window, $PerturbType$ -
benign or random perturbation, 
\par\end{raggedright}
\noindent \begin{raggedright}
$IterationMethod$ - logarithmic backtracking or linear iteration,
$AttackerKnowledge$ - decision or score-based.
\par\end{raggedright}
\LyXZeroWidthSpace{}

\textbf{for} each sliding windows \textbf{$(\boldsymbol{w_{j}^{m}},\boldsymbol{w_{j}^{b}}$)}
of $n$ API calls in $(\boldsymbol{x^{m}},\boldsymbol{x^{b}})$, respectively:

\ \ \ \ $\boldsymbol{(w_{j}^{m}},addedAPIs)=Algorithm2(f,\boldsymbol{w_{j}^{m}},\boldsymbol{w_{j}^{b}},n,D^{'},M_{w},PerturbType,IterationMethod,AttackerKnowledge)$\}

\ \ \ \ \textbf{if} $IterationMethod$ is logarithmic backtracking:

\ \ \ \ \ \ \ \ $remainingAPIs=addedAPIs$

\ \ \ \ \ \ \ \ \textbf{while} $(f(\boldsymbol{w_{j}^{m}})=benign)$:

\ \ \ \ \ \ \ \ \ \ \ \ \# Remove added API calls until
evasion is lost:

\ \ \ \ \ \ \ \ \ \ \ \ Randomly split $addedAPIs$ into
two equally sized groups: $remainingAPIs,deletedAPIs$

\ \ \ \ \ \ \ \ \ \ \ \ remove $deletedAPIs$ from $\boldsymbol{w_{j}^{m}}$

\ \ \ \ \ \ \ \ \ \ \ \ \textbf{if} $f(\boldsymbol{w_{j}^{m}})=malicious:$

\ \ \ \ \ \ \ \ \ \ \ \ \ \ \ \ $\boldsymbol{w_{j}^{m}=w_{j}^{m}}+deletedAPIs-remainingAPIs$
\# Remove $remainingAPIs$ instead of $deletedAPIs$ from $\boldsymbol{w_{j}^{m}}$

\ \ \ \ \ \ \ \ \ \ \ \ \ \ \ \ Switch between $remainingAPIs$
and $deletedAPIs$

\ \ \ \ \ \ \ \ $recoveredAPIs=deletedAPIs$

\ \ \ \ \ \ \ \ \textbf{while} $(f(\boldsymbol{w_{j}^{m}})=malicious)$:

\ \ \ \ \ \ \ \ \ \ \ \ \# While there are still added
API calls that were removed, add them back until evasion is restored:

\ \ \ \ \ \ \ \ \ \ \ \ $recoveredAPIs=$Randomly pick
half of the API calls remaining in $deletedAPIs$

\ \ \ \ \ \ \ \ \ \ \ \ Add $recoveredAPIs$ to $\boldsymbol{w_{j}^{m}}$

\textbf{return} (perturbed) $\boldsymbol{x^{m}}$
\end{algorithm*}

In Algorithm \ref{alg:Random-Window-Sequence}, the attacker chooses
the API calls to add and remove randomly. Note that we do not remove
the malware's original API calls (only the no-op API calls that were
previously added by the adversary), in order to prevent harm to its
functionality. Since we add or remove half of the API calls each time,
we perform $O(\log n)$ queries per adversarial sliding window if
$IterationMethod$ is logarithmic backtracking, instead of the $O(n)$
queries that are performed if $IterationMethod$ is linear iteration
(where $n$ is the size of the adversarial sliding window), making
this attack query-efficient, as can be seen in Tables \ref{tab:Attack-Comparison}
and \ref{tab:Benign-Perturbation-Attack-1}. While the proposed attack
is designed for API call-based classifiers, it can be used for any
adversarial sequence generation. 

The benign sequence \textbf{$\boldsymbol{x^{b}}$} is generated by
a specially crafted GAN, which is described below.

\subsubsection{\label{subsec:Benign-Perturbation:-GAN-1}Perturbation Type}

In this subsection, we describe the methods to select what API calls
to add by our attack. As an alternative to choosing the API calls
randomly from all available API calls (\emph{random perturbation}),
we propose the \emph{benign perturbation} method. In this method,
instead of adding random API calls, we add API calls selected from
sequences generated by a generative adversarial network (GAN) that
has been trained to mimic real benign sequences. This concept is inspired
by the modus operandi of biological viruses (malware) which are sometimes
composed of human (``benign'') proteins in order to evade the immune
system (malware classifier) of the host.

When the attackers add an API call to our adversarial sequence, they
want to have the maximum impact on the classifier's score. Thus, Algorithm
\ref{alg:Logarithmic-Backtracking-Attack} can take $\boldsymbol{x^{b}}$,
a benign API sequence, as input. The idea is that adding a ``benign''
API call would make the trace ``more benign'' than adding a random
API call. This is due to the fact that no API call is malicious or
benign per se. The context and flow of API calls determine the functionality
(and therefore the code's ``maliciousness''). Using benign perturbation
creates a ``benign API call context'' and thus improves the attack
effectiveness and also makes it more query-efficient, because fewer
queries are needed to fool the classifier, as can be seen in Table
\ref{tab:Benign-Perturbation-Attack-1}, which uses benign perturbations,
as opposed to Table \ref{tab:Attack-Comparison}, which uses linear
iteration. 

One way to generate \textbf{$\boldsymbol{x^{b}}$} is by taking the
API call sequence of an actual benign sample from our dataset. The
downside of this approach is that those hard-coded API calls can be
detected explicitly as an evasion attack. 

A better approach is to generate a different benign sequence each
time, using a generative model. One way to do this is to use a generative
adversarial network (GAN), with a stochastic input seed and an output
of an API call sequence that is indistinguishable (to the discriminator
classifier) from actual benign sequences from the dataset. This approach
is rarely used for API call sequence generation, but it has been used
for text generation. Note that this approach does not require queries
to the target classifier (unlike, e.g., building a substitute model,
as done in \citep{DBLP:conf/raid/RosenbergSRE18}). 

In comparison to other approaches (e.g., VAE), using a GAN tends to
generate better output. Most other methods require that the generative
model has some particular functional form (like a Gaussian output
layer). Moreover, all of the other approaches require that the generative
model puts non-zero mass everywhere. 

However, a challenge with the GAN approach is that the discrete outputs
from the generative model make it difficult to pass the gradient update
from the discriminative model to the generative model. Another challenge
is that the discriminative model can only classify a complete input
sequence. We used SeqGAN \citep{DBLP:conf/aaai/YuZWY17} implementation.
Following a pretraining procedure that follows the MLE (maximum likelihood
estimation) metric, the generator $G$ is modeled as a stochastic
policy in reinforcement learning (RL). The agent's state is the API
call subsequence of the first $t$ API call types in the sequence
to be generated by the GAN, $s_{t}=\left[x_{0},x_{1},..x_{t-1}\right]$.
The agent's action is the next API call type in the sequence to be
generated by the SeqGAN model $x_{t}\sim\left(x|s_{t}\right)$. The
reward is the feedback given to $G$ by $D$ when evaluating the generated
sequence, bypassing the gradient update challenge by directly performing
a gradient policy update. 

In a stochastic parameterized policy, the actions are drawn from a
distribution that parameterizes the policy. An action may be sampled
from, e.g., a normal distribution whose mean and variance will be
predicted by the policy. The objective of $G$ is to generate a sequence
from the start state $s_{0}$ in such a way that maximizes the expected
end reward. This action-value function is estimated by the discriminator.
However, $D$ only provides a reward at the end of a finished sequence.
Yet, it is important that at every time-step, the fitness of both
previous tokens as well as future outcome are considered. For this,
the policy gradient used in SeqGANs employs a Monte Carlo search with
a roll-out policy (P) to sample the unknown remaining tokens and approximates
the state-action value in an intermediate step. 

We trained our GAN using a benign hold-out set of 3,000 sequences
that were taken from the same distribution as the test set, available
to the attacker. This hold-out set was not used subsequently as part
of the test set, to avoid data leak to the attacker, artificially
increasing the attack effectiveness.

We also tried other GAN architectures (e.g., GSGAN), but SeqGAN outperformed
all of them (Appendix \ref{sec:Appendix-C:-Benign}). SeqGAN outperformed
the random perturbation as well, as shown in Section \ref{subsec:Attack-Performance}.

\subsubsection{\label{subsec:Decision-Based-Attack}\label{subsec:Score-Based-Query-Efficient}Attacker
Knowledge}

This characteristic entails a trade off: the more information the
attackers have about the classifier, the more query-efficient their
attack would be. In this paper, the attacker may have access to the
confidence score of the malware classifier (\emph{score-based attack}),
or only to the predicted label (\emph{decision-based attack}). For
the first case, we propose a score-based attack that uses a gradient-free
optimization algorithm that until now has never been used for adversarial
learning, outperforming state-of-the-art attacks for low number of
queries to the attacked classifier. This attack is designed for discrete
values in sequences of variable length. Thus, it fits API call sequences,
as opposed to image pixels, which were the focus of most previous
research.

While some malware classifiers do expose their confidence score (e.g.,
MAX, a non-sequence based NGAV product in VirusTotal online scanning
service \citep{VirusTotal}), others do not. Therefore, we also implemented
a decision-based attack, which is less query-efficient, but requires
only knowledge about the predicted label of the malware classifier.

\paragraph{Decision-Based Attack}

In Algorithm \ref{alg:Random-Window-Sequence}, we show how to generate
an adversarial sequence for a single API call window (out of the entire
API call sequence).

\begin{algorithm*}
\caption{\label{alg:Random-Window-Sequence}Single Iteration Decision-Based
Window Sequence Generation}

\textbf{Input}: $f$ - black-box model, \textbf{$\boldsymbol{w^{m}}$}
- malicious sequence to perturb, of length $l^{m}\leq n$, 

\textbf{$\boldsymbol{w^{b}}$} - benign sequence to mimic, of length
$l^{b}\leq n$, $n$ - size of adversarial sliding window, 

$D^{'}$ - adversarial vocabulary, $M_{w}$ - maximum API modifications
per window, $PerturbType$ - benign or random perturbation,

$IterationMethod$ (logarithmic backtracking or linear iteration.

\LyXZeroWidthSpace{}

$addedAPIs=\{\}$

\textbf{while} (($IterationMethod$ is linear iteration) and $(f(\boldsymbol{w^{m}})=malicious)$)
or $(|addedAPIs|<M_{w})$: 

\ \ \ \ Randomly select an API's position $i$ in $\boldsymbol{w^{m}}$

\ \ \ \ \textbf{if} $PerturbType$ is benign perturbation:

\ \ \ \ \ \ \ \ Add $\boldsymbol{w^{b}}[i]$ to $\boldsymbol{w^{m}}$
at position $i$ 

\ \ \ \ \textbf{else:} $PerturbType$ is random perturbation

\ \ \ \ \ \ \ \ Add a random API in $D^{'}$ to $\boldsymbol{w^{m}}$
in position $i$ 

\ \ \ \ Add the new API and its position to $addedAPIs$

\textbf{if} $(f(\boldsymbol{w^{m}})=malicious)$ and $(|addedAPIs|=M_{w})$:
\textbf{return} Failure

\textbf{return} $\boldsymbol{(w^{m}},addedAPIs)$ \# \textbf{$\boldsymbol{w^{m}}$}
includes the perturbation
\end{algorithm*}

The perturbation added is either random API calls, a.k.a. \emph{random
perturbation} (line 12), or API calls of a benign sequence, a.k.a.
\emph{benign perturbation} (line 10). Since only the predicted class
is available, the API calls are added in a random position $i$ in
the API sequence (line 8). The adversaries randomly chooses $i$,
since they do not have a better way of selecting $i$ without incurring
significant statistical overhead. Note that the addition of an API
call in position $i$ means that the API calls from position $i..n$
($\boldsymbol{w^{m}}[i..n]$ ) are ``pushed back'' one position
to make room for the new API call, in order to maintain the original
sequence and preserve the original functionality of the code (in line
10 and 12). Since the sliding window has a fixed length, the last
API call, $\boldsymbol{w^{m}}[n+1]$, is ``pushed out'' and removed
from $\boldsymbol{w^{m}}$. This API call addition continues until
the modified sequence $\boldsymbol{w^{m}}$ is classified as benign
or more than $M_{w}$ API calls are added, reaching the maximum overhead
limit (line 7). In this case the attack has failed (line 14). In the
case of a \emph{linear iteration} attack, the API calls are added
one at a time, checking whether the perturbed sequence evades detection
after each addition. The case of a \emph{logarithmic backtracking}
attack was explained in Section \ref{subsec:Full-Sequence-Linear}.

One might claim that a simpler attack can be used instead: insert
$n-1$ no-op API calls after each API call from the original binary.
This attack effectiveness is 100\%, and no queries are needed to implement
it, making it extremely query-efficient. However, this trivial attack
has two major issues:
\begin{enumerate}
\item It is easy to detect this trivial attack using anomaly detection,
since no actual benign program call trace is composed like that.
\item Such malware would run much slower than the original malware due to
the additional API overhead, allowing the intrusion prevention systems
of the victim to mitigate such malware, e.g., by terminating perturbed
ransomware, after encrypting only several files, due to its perturbation
induced slowness.
\end{enumerate}

\paragraph{Score-Based Attack}

When the confidence score of the malware classifier is also returned,
score-based attacks (e.g., gradient-free optimization algorithms)
can be applied as well. The merged flow for both attacks is described
in Algorithm \ref{alg:Full-Window-Sequence}. Assuming the attacker
has a budget of $B$ queries per API call window, the call to Algorithm
\ref{alg:Random-Window-Sequence} in line 9 of Algorithm \ref{alg:Full-Window-Sequence}
can be replaced with $B-\log n$ iterations (line 11) of minimizing
$f(\boldsymbol{w^{m}})$ (lines 13 and 15) by one of the score minimization
algorithms presented in Section \ref{subsec:Comparison-to-Previous}.
In order to use the same budget for all attacks, we chose $B=M_{w}$
(lines 9, 11). 

All random perturbation variants try to minimize the target classifier
score by modifying only the values of the added API call types (while
the API call positions are random but fixed, as in Algorithm \ref{alg:Random-Window-Sequence}).
Trying to modify both API types and positions with the same budget
results in inferior performance (this is not shown due to space limits).
All benign perturbation variants try to minimize the score by modifying
only the API positions (while the API types are taken from the GAN's
output). 

The maximum additional API calls allowed per sliding window was set
to 70 (50\%). The search space for this optimization would either
be the $M_{w}$ added API call type values (out of $|D|$ values each)
if this is a random perturbation, or the $M_{w}$ added API call indices
in the adversarial window (each with $n$ possible values) if this
is a benign perturbation. 

\paragraph{Score-Based Query-Efficient Attack for Discrete Input Sequence}

Most state-of-the-art query-efficient attacks for images assume continuous
input (\citep{DBLP:conf/icml/UesatoOKO18,DBLP:conf/icml/IlyasEAL18})
and underperform when used for discrete input (e.g., API calls or
position indices), as shown in Table \ref{tab:Attack-Comparison}.
Genetic algorithms (GAs), which use mutation (random perturbation)
in existing adversarial candidates and crossover between several candidates
(i.e., a combination of parts of several candidates), are an exception.
GAs work well with discrete sequences \citep{DBLP:conf/emnlp/AlzantotSEHSC18}.
However, while crossover makes sense in the NLP domain (e.g., for
compound sentences), it makes little sense for API call sequences,
where each program has its own business logic. Another issue is the
poor performance of a fixed mutation rate, usually used by GAs. It
is better to use an adaptive mutation rate, which fits itself to the
domain without knowledge expertise \citep{10.1007/978-3-319-45823-6_75}.

We decided to use the self-adaptive uniform mixing evolutionary algorithm
(EA) proposed by Dang et al. \citep{10.1007/978-3-319-45823-6_75}.
It starts with a population of adversarial candidates, and in every
generation, a new population of candidates is produced, and the old
generation dies. Besides the adversarial sequence, each candidate
carries an additional property: its mutation rate. Each new candidate
is produced in the same way:
\begin{enumerate}
\item The best of two uniformly selected individuals is selected (i.e.,
tournament selection).
\item The selected parent individual changes its mutation rate between two
mutation rates: low and high, with a fixed probability $p$. We used
the values proposed in \citep{10.1007/978-3-319-45823-6_75}.
\item The parent replicates, with mutations occurring at the new mutation
rate.
\end{enumerate}
Although the selection mechanism does not take into account the mutation
rate, the intuition is that appropriate mutation rates are correlated
with high fitness.

We assume that the EA attack would be query-efficient due to the following
reasons:
\begin{enumerate}
\item The usage of the adaptive mutation rate helps reducing the number
of queries, before the highest impact element is added to the sequence.
\item Unlike other attacks (including other self-adaptive heuristic strategies,
e.g., \citep{DBLP:conf/icml/UesatoOKO18}), EA is effective in a discrete
feature space. This makes it more query-efficient, because fewer queries
are needed before it fools the target classifier, as can be seen in
Tables \ref{tab:Attack-Comparison} and \ref{tab:Benign-Perturbation-Attack-1}.
\item The crossover (combining API calls from malicious and benign parents,
instead of mutating a malicious parent) used by other attacks (e.g.,
GA, also efficient in discrete feature spaces) makes no sense in the
cyber domain, because it adds redundant API calls of the benign parent.
Thus, an algorithm that forgoes the crossover (and thus the addition
of redundant API calls) is more query-efficient in the cyber domain.
\end{enumerate}
\begin{algorithm*}
\caption{\label{alg:Full-Window-Sequence}Score-Based or Decision-Based Window
Sequence Generation}

\textbf{Input}: $f$ - black-box model, 

\textbf{$\boldsymbol{w^{m}}$} - malicious sequence to perturb, of
length $l^{m}\leq n$, \textbf{$\boldsymbol{w^{b}}$} - benign sequence
to mimic, of length $l^{b}\leq n$, 

$n$ - size of adversarial sliding window, $D^{'}$ - adversarial
vocabulary, 

$M_{w}$ - maximum API modifications per window, $PerturbType$ -
benign or random perturbation, 

$IterationMethod$ - logarithmic backtracking or linear iteration,
$AttackerKnowledge$ - decision or score-based.

\LyXZeroWidthSpace{}

\textbf{if} $AttackerKnowledge$ is decision-based:

\ \ \ \ \textbf{while} $(f(\boldsymbol{w^{m}})=malicious):$

\ \ \ \ \ \ \ \ $\boldsymbol{(w^{m}},addedAPIs)=Algorithm2(f,\boldsymbol{w^{m}},\boldsymbol{w^{b}},n,D^{'},M_{w},PerturbType,IterationMethod)$\}

\textbf{else}: \#$AttackerKnowledge$ is score-based

\ \ \ \ \textbf{if} $IterationMethod$ is logarithmic backtracking
$optimIterationsCount=M_{w}-\lg n$ , else $optimIterationsCount=M_{w}$

\ \ \ \ \textbf{if} $PerturbType$ is benign perturbation:

\ \ \ \ \ \ \ \ $\boldsymbol{(w^{m}},addedAPIs)=ScoreMinimizationAlgorithm$($f(\boldsymbol{w^{m}}),optimIterationsCount,addedAPIIndices$)

\ \ \ \ \textbf{else}: \# $PerturbType$ is random perturbation:

\ \ \ \ \ \ \ \ $\boldsymbol{(w^{m}},addedAPIs)=ScoreMinimizationAlgorithm$($f(\boldsymbol{w^{m}}),optimIterationsCount$,$addedAPIValues$)

\textbf{return} $\boldsymbol{(w^{m}},addedAPIs)$ \# \textbf{$\boldsymbol{w^{m}}$}
includes the perturbation
\end{algorithm*}

\section{Experimental Evaluation}

\subsection{\label{subsec:Dataset-and-Target}Dataset and Target Malware Classifiers}

We use the same dataset used in \citep{DBLP:conf/raid/RosenbergSRE18},
because of its size: it contains 500,000 files (250,000 benign samples
and 250,000 malware samples), faithfully representing the malware
families in the wild and providing a proper setting for an attack
comparison. Details about the dataset are provided in Appendix \ref{sec:Appendix-A:-Tested}. 

Each sample was run in Cuckoo Sandbox, a malware analysis system,
for two minutes per sample. The API call sequences generated by the
inspected code were extracted from the JSON report generated by Cuckoo
Sandbox. The extracted API call sequences are used as the malware
classifier's features. The samples were run on dozens of Windows 8.1
OS instances on the cloud, since most malware targets the Windows
OS. Anti-sandbox malware was filtered to prevent dataset contamination
(see Appendix \ref{sec:Appendix-A:-Tested}). After filtering, the
final training set size is 360,000 samples, 36,000 of which serve
as the validation set. The test set size is 36,000 samples. All sets
are balanced between malicious and benign samples. 

While some ML-based dynamic analysis cloud services are used by enterprises,
e.g., \citep{JoeSandboxML}, there are no trial versions of commercial
products or open source API call-based deep learning malware classifiers
available (such commercial products target enterprises and involve
supervised server installation). Dynamic models are also unavailable
on VirusTotal. In order to compensate for this, we used the malware
classifiers detailed in Appendix \ref{sec:Appendix-B:-Tested} and
simulated the cloud service use case by deploying Keras \citep{Keras}
models on Amazon cloud using SageMaker \citep{DeployKerasModelAmazon},
and then we queried them by accessing the cloud service. 

The API call sequences are split into windows of $k$ API calls each,
and each window is classified in turn. Thus, the input of each of
the classifiers is a vector of $k=140$ (larger window sizes, such
as $k=1,000$, didn't improve the classifier's accuracy) API call
types with 314 possible values (those monitored by Cuckoo Sandbox,
mentioned in \citep{CuckooHookedAPIs}). 

The implementation and hyperparameters (loss function, dropout, activation
functions, etc.) of the target classifiers are described in Appendix
\ref{sec:Appendix-B:-Tested}. 

On the test set, all of the DNN classifiers achieve over 95\% accuracy,
and all other classifiers reach over 90\% accuracy. The classifiers'
false positive rate ranged from 0.5 to 1\%.

\subsection{\label{subsec:Attack-Performance}Attack Performance}

\subsubsection{\label{subsec:Attack-Performance-Metrics}Attack Performance Metrics}

In order to measure the performance of an attack, we consider three
factors (by a descending order of importance):

We consider the average number of malware classifier queries the attack
performs per adversarial example before it is classified as benign
by the malware classifier. The attacker aims to minimize this number,
since in cloud scenarios, each query costs money and increases the
probability of adversarial attempt detection.

We also consider the \emph{attack effectiveness}, which is the percentage
of malicious samples which were correctly classified by the malware
classifier for which the adversarial sequence $\boldsymbol{x^{m}}^{*}$
generated by Algorithm \ref{alg:Logarithmic-Backtracking-Attack}
was misclassified as benign by the malware classifier. An attack is
defined as query-efficient if it has the highest attack effectiveness
for a given (fixed) number of queries. (A different approach of a
fixed accuracy is computationally expensive to compute.)

Finally, we consider the \emph{attack overhead}, that is, the fraction
of API calls which were added (by Algorithm \ref{alg:Logarithmic-Backtracking-Attack})
to the malware samples, out of the total number of API calls.

The average length of the API call sequence is: $avg(length(\boldsymbol{x^{m}}))\approx10,000$.
We used a maximum of $M_{w}=70$ additional API calls per window of
$k=140$ API calls, limiting the perturbation run time overhead (per
window and thus per sample) to 50\% in the worst case. While not shown
here due to space limits, higher $M_{w}$ values cause higher average
attack effectiveness and overhead, and more queries.

Adversarial attacks against images usually try to minimize the number
of modified pixels in order to evade human detection of the perturbation.
One might claim that such definition of minimal perturbation is irrelevant
for API call traces: humans cannot inspect sequences of thousands
or millions of APIs, so an attacker can add an unlimited amount of
API calls. However, one should bear in mind that a malware aims to
perform its malicious functionality as fast as possible. For instance,
ransomware usually starts by encrypting the most critical files (e.g.,
the 'My Documents' folder) first, so even if the user turns off the
computer and sends the hard-drive to IT - damage has already been
done. The same is true for a key-logger - it aims to send the user
passwords to the attacker as soon as possible, so they can be used
immediately, before the malware is detected and neutralized. Moreover,
adding too many API calls would cause the modified program's profile
to become anomalous, making it easier for anomaly detection intrusion
detection systems, e.g., systems that measure CPU usage \citep{DBLP:conf/isi/MoskovitchPGSFPSE07}
or contain irregular API call subsequences \citep{Grosse2017a} to
detect anomalies.

\subsubsection{\label{subsec:Comparison-to-Previous}Comparison to Previous Work}

\paragraph{Decision-Based Attack Performance}

A comparison of the attack effectiveness and attack overhead of our
decision-based attack with logarithmic backtracking transformation
and with benign perturbation (Algorithm \ref{alg:Logarithmic-Backtracking-Attack})
to BiRand, the more efficient attack used in Dang et al. \citep{DBLP:conf/ccs/DangHC17},
and to Rosenberg et al. for different attacked malware classifiers
is presented in Table \ref{tab:Attack-Performance-of} (an average
of five runs, with 100 queries).

\begin{table*}[t]
\caption{\label{tab:Attack-Performance-of}Decision-Based Attack Performance
(100 Queries)}

\centering{}%
\begin{tabular}{|>{\centering}p{0.2\textwidth}|>{\centering}p{0.12\textwidth}|>{\centering}p{0.1\textwidth}|>{\centering}p{0.1\textwidth}|>{\centering}p{0.12\textwidth}|>{\centering}p{0.1\textwidth}|>{\centering}p{0.1\textwidth}|}
\hline 
Classifier Type & Attack Effectiveness {[}\%{]}

Our Decision-Based Attack & Attack Effectiveness {[}\%{]}

BiRand, Dang et al. 2017 \citep{DBLP:conf/ccs/DangHC17} & Attack Effectiveness {[}\%{]}

Rosenberg et al. 2018 \citep{DBLP:conf/raid/RosenbergSRE18} & Additional API Calls {[}\%{]}

Our Decision-Based Attack & Additional API Calls {[}\%{]}

BiRand, Dang et al. 2017 \citep{DBLP:conf/ccs/DangHC17} & Additional API Calls {[}\%{]}

Rosenberg et al. 2018 \citep{DBLP:conf/raid/RosenbergSRE18}\tabularnewline
\hline 
\hline 
LSTM & \textbf{62.21} & 39.49 & 51.15 & 22.22 & 31.42 & 17.22\tabularnewline
\hline 
\hline 
Deep LSTM & \textbf{63.62} & 40.38 & 50.80 & 22.71 & 32.12 & 29.51\tabularnewline
\hline 
\hline 
GRU & \textbf{63.35} & 40.21 & 51.16 & 21.47 & 30.36 & 16.09\tabularnewline
\hline 
\hline 
1D CNN & \textbf{63.63} & 40.39 & 48.93 & 4.10  & 5.80 & 49.21\tabularnewline
\hline 
\hline 
Logistic Regression & \textbf{41.47} & 26.32 & 35.67 & 4.43  & 6.26 & 7.58\tabularnewline
\hline 
\hline 
Random Forest & \textbf{63.24} & 40.14 & 50.87 & 5.20  & 7.35 & 9.40\tabularnewline
\hline 
\hline 
SVM & \textbf{42.59} & 27.04 & 36.27 & 3.82 & 5.40 & 7.19\tabularnewline
\hline 
\hline 
Gradient Boosted Tree & \textbf{41.62} & 26.41 & 36.55 & 13.99 & 19.78 & 27.80\tabularnewline
\hline 
\end{tabular}
\end{table*}

Rosenberg et al. provides state-of-the-art performance for gradient-based
attacks against a wide range of classifiers. Dang et al. \citep{DBLP:conf/ccs/DangHC17}
presented a decision-based attack similar to ours, but limited to
non-sequential features. The attack effectiveness of our logarithmic
backtracking attack, shown in Table \ref{tab:Attack-Comparison} (logarithmic
backtracking=yes columns), is identical to the BiRand algorithm presented
in \citep{DBLP:conf/ccs/DangHC17} (because both attacks use a similar
algorithm).

As can be seen in Table \ref{tab:Attack-Performance-of}, our proposed
decision-based attack has the highest effectiveness for all of the
malware classifiers tested. Rosenberg et al. provides higher attack
effectiveness than our decision-based attack for 2500 queries (not
shown due to space limits), but underperforms when the number of queries
is being reduced, because there aren't enough queries to build a substitute
model with accurate decision boundary. 

We see that our attack provides higher attack effectiveness than Dang
et al., due to our attack's use of benign perturbation, which was
not presented in \citep{DBLP:conf/ccs/DangHC17}. When modifying BiRand
to use benign perturbation, the results are identical to those obtained
by our attack (because both attacks use a similar algorithm; this
is not shown due to space limits). In addition, we see that our attack
produces a smaller perturbation (adding 25-50\% less API calls; see
Table \ref{tab:Attack-Performance-of}) than BiRand. This is due to
the additional level of backtracking in our attack (lines 19-22 in
Algorithm \ref{alg:Logarithmic-Backtracking-Attack}). This backtracking
is not available in BiRand, which implements a binary search. 

As mentioned in Section \ref{subsec:Dataset-and-Target}, $|TestSet(f)|=36,000$
samples, and the test set $TestSet(f)$ is balanced, so the attack
performance was measured on: $|\{f(\boldsymbol{x_{m}})=Malicious|\boldsymbol{x_{m}}\in TestSet(f)\}|=18,000$
samples.

We used $k=n$ for Algorithm \ref{alg:Random-Window-Sequence}, i.e.,
the non overlapping sliding window size of the adversary is the same
as that used by the target classifier. However, even if this is not
the case, the attack effectiveness is not significantly degraded.
If $n<k$, the adversary can only modify a subset of the API calls
affecting the target classifier, and this subset might not be diverse
enough to affect the classification as desired, thereby reducing the
attack effectiveness. If $n>k$, the adversaries would keep trying
to modify different API calls' positions in Algorithm \ref{alg:Random-Window-Sequence},
until they modify the ones impacting the target classifier as well,
thereby increasing the attack overhead without affecting the attack
effectiveness. For instance, when $n=100,\,k=140$, there is an average
decrease in attack effectiveness from 87.96\% to 87.94\% for an LSTM
classifier. Other classifiers have similar behavior, which is not
shown due to space limits. The closer $n$ and $k$ are, the better
the attack performance. 

We used the adversarial vocabulary:\\
 $D'=D-\{ExitWindowsEx(),NtTerminateProcess()\}$, where $D$ is all
of the API call types recorded by the malware classifier, so $D'$
does not contain any API type that might harm the code's functionality. 

\paragraph{Score-Based Attack Performance}

There are no published query-efficient adversarial attacks against
RNN variants. Attacks that minimize the number of queries exist, but
they only work against CNNs \citep{DBLP:conf/icml/IlyasEAL18,DBLP:journals/corr/abs-1805-11090}.
Those attacks aren't relevant, because they don't work with sequence
input and discrete values. To address this gap, we used Nevergrad
\citep{nevergrad}, a gradient-free optimization library, to implement
discrete sequence input variants of a few state-of-the-art score-based
adversarial attacks:
\begin{enumerate}
\item SPSA-based attack (Uesato et al. \citep{DBLP:conf/icml/UesatoOKO18}).
\item NES-based attack (Ilyas et al. \citep{DBLP:conf/icml/IlyasEAL18}).
\item GA-based attack (Alzantot et al. \citep{DBLP:conf/emnlp/AlzantotSEHSC18},
Xu et al. \citep{DBLP:conf/ndss/XuQE16}).
\item The gradient-based attack (Rosenberg et al. \citep{DBLP:conf/raid/RosenbergSRE18}).
\end{enumerate}
We compare theses attacks to our score-based uniform mixing EA attack
(described in Section \ref{subsec:Score-Based-Query-Efficient}) and
to our decision-based attack. We didn't implement the ZOO attack of
Chen et al. \citep{Chen2017}, because it has already been evaluated
and was found to be less effective than both SPSA and NES attacks
\citep{DBLP:conf/icml/UesatoOKO18}.

We used Nevergrad's default arguments for all attacks. 

The attack performance (average of five runs) for the LSTM classifier
with a fixed budget of 100, 200, and 2500 queries (the attack of Rosenberg
et al. requires many queries to accurately build the substitute model
required to estimate the gradients per API window \citep{DBLP:conf/raid/RosenbergSRE18})
is presented in Table \ref{tab:Attack-Comparison} for random \emph{perturbation
type} attacks and in Table \ref{tab:Benign-Perturbation-Attack-1}
for benign \emph{perturbation type} attacks (Rosenberg et al. has
the same performance in both tables because its perturbations are
always determined by the maximum gradient).

The first two lines of each table pertain to our attacks with different
\emph{attacker knowledge}. When combining these lines with the \emph{iteration
method} values in Tables \ref{tab:Attack-Comparison} and \ref{tab:Benign-Perturbation-Attack-1},
one gets our eight previously described attacks (all combinations
of: \emph{iteration method}, \emph{attacker knowledge} and \emph{perturbation
type}). The \emph{iteration method} values are Linear (for \emph{linear
iteration}) and Log (for \emph{logarithmic backtracking}). Other classifiers
and budgets (not shown due to space limits) resulted in similar relative
trends: a higher budget results in increased attack effectiveness.
Note that here we use a fixed number of queries and try to maximize
the attack effectiveness for the specified number of queries, since
the reverse approach requires higher computational effort and yields
the same results. 

\begin{table*}[t]
\caption{\label{tab:Attack-Comparison}Random \emph{Perturbation Type} Attack
Effectiveness Comparison for a Fixed Number of Queries (LSTM Model)}

\centering{}%
\begin{tabular}{|>{\centering}p{0.4\textwidth}||>{\centering}p{0.07\textwidth}|>{\centering}p{0.07\textwidth}|>{\centering}p{0.07\textwidth}|>{\centering}p{0.07\textwidth}|>{\centering}p{0.07\textwidth}|>{\centering}p{0.07\textwidth}|}
\hline 
Number of Queries & 100 & 200 & 2500 & 100 & 200 & 2500\tabularnewline
Logarithmic Backtracking (/BiRand)\emph{Iteration Method} & Linear & Linear & Linear & Log & Log & Log\tabularnewline
\hline 
\hline 
Our Score-Based Attack (Score-Based \emph{Attacker Knowledge}) & \textbf{58.75} & \textbf{67.59} & \textbf{100.00} & \textbf{69.28} & \textbf{79.71} & \textbf{100.00}\tabularnewline
\hline 
\hline 
Our Decision-Based Attack (Decision-Based \emph{Attacker Knowledge}) & 19.86 & 21.25 & 31.43 & 39.49 & 42.25 & 62.50\tabularnewline
\hline 
\hline 
Rosenberg et al. \citep{DBLP:conf/raid/RosenbergSRE18} & 51.15 & 67.11 & 99.99 & 51.15 & 67.11 & 99.99\tabularnewline
\hline 
\hline 
Uesato et al. \citep{DBLP:conf/icml/UesatoOKO18} & 2.37 & 2.73 & 2.87 & 5.17 & 5.95 & 6.25\tabularnewline
\hline 
\hline 
Ilyas et al. \citep{DBLP:conf/icml/IlyasEAL18} & 37.50 & 43.14 & 74.94 & 43.78 & 50.37 & 87.50\tabularnewline
\hline 
\hline 
Alzantot et al. \citep{DBLP:conf/emnlp/AlzantotSEHSC18}, Xu et al.
\citep{DBLP:conf/ndss/XuQE16} & 54.68 & 62.91 & 100.00 & 62.06 & 71.40 & 100.00\tabularnewline
\hline 
\end{tabular}
\end{table*}

\begin{table*}
\caption{\label{tab:Benign-Perturbation-Attack-1}Benign \emph{Perturbation
Type} Attack Effectiveness Comparison for a Fixed Number of Queries
(LSTM Model)}

\centering{}%
\begin{tabular}{|>{\centering}p{0.4\textwidth}||>{\centering}p{0.07\textwidth}|>{\centering}p{0.07\textwidth}|>{\centering}p{0.07\textwidth}|>{\centering}p{0.07\textwidth}|>{\centering}p{0.07\textwidth}|>{\centering}p{0.07\textwidth}|}
\hline 
Number of Queries & 100 & 200 & 2500 & 100 & 200 & 2500\tabularnewline
Logarithmic Backtracking (/BiRand)\emph{Iteration Method} & Linear & Linear & Linear & Log & Log & Log\tabularnewline
\hline 
\hline 
Our Score-Based Attack (Score-Based \emph{Attacker Knowledge}) & \textbf{71.90} & \textbf{82.70} & \textbf{100.00} & \textbf{84.77} & \textbf{97.53} & \textbf{100.00}\tabularnewline
\hline 
\hline 
Our Decision-Based Attack (Decision-Based \emph{Attacker Knowledge}) & 41.34 & 44.24 & 46.34 & 62.21 & 63.97 & 87.96\tabularnewline
\hline 
\hline 
Rosenberg et al. \citep{DBLP:conf/raid/RosenbergSRE18} & 51.15 & 67.11 & 99.99 & 51.15 & 67.11 & 99.99\tabularnewline
\hline 
\hline 
Uesato et al. \citep{DBLP:conf/icml/UesatoOKO18} & 6.56 & 7.55 & 11.47 & 14.31 & 16.46 & 25.00\tabularnewline
\hline 
\hline 
Ilyas et al. \citep{DBLP:conf/icml/IlyasEAL18} & 66.23 & 76.19 & 81.25 & 77.32 & 88.96 & 94.87\tabularnewline
\hline 
\hline 
Alzantot et al. \citep{DBLP:conf/emnlp/AlzantotSEHSC18}, Xu et al.
\citep{DBLP:conf/ndss/XuQE16} & 68.49 & 81.09 & 100.00 & 79.93 & 91.96 & 100.00\tabularnewline
\hline 
\end{tabular}
\end{table*}

Our score-based attack variants (in Tables \ref{tab:Attack-Comparison}
and \ref{tab:Benign-Perturbation-Attack-1}) provide a higher attack
effectiveness for all classifiers (this is not discussed further due
to space limits) because of the more efficient search algorithms used.
The attack of Rosenberg et al. requires using 2255 queries per API
call window to build the substitute model accurately enough to reach
the performance mentioned in \citep{DBLP:conf/raid/RosenbergSRE18}
(the rest of the queries, for a a total of 2500, are required to perform
the attack itself). Trying to use fewer queries results in a substitute
model with inaccurate decision boundary that affect the gradients,
and thus the gradient based attack effectiveness. 

One might claim that the same substitute model can be used to camouflage
more than ten malware samples, resulting in a lower average budget
per sample. However, in most cases an attacker would try to modify
only a single malware, so it can bypass the detector and perform its
malicious functionality. Moreover, even if the average cost per example
can be reduced by using the same substitute model, our attack presents
a lower limit on the \emph{absolute} number of queries, bypassing
a cloud-service that blocks access for a host performing too many
queries in a short amount of time in order to thwart adversarial efforts
\citep{chen2019stateful,chau2010polonium}. The more efficient the
attack, the less chances there are for it to be mitigated by this
approach. 

The performance of our linear iteration attack, shown in Table \ref{tab:Attack-Comparison}
(logarithmic backtracking=no columns), is identical to the performance
of the SeqRand algorithm presented in \citep{DBLP:conf/ccs/DangHC17}
(because both attacks use the same algorithm).

The attack overhead of all attacks is similar: about 30\%, or 40 API
calls, per window. Since a classifier with an API window size of $k=100$
provides roughly the same accuracy as with $k=140$ used here (96.76\%
vs. 97.61\% with the same FP rate for the LSTM classifier), the success
of these attacks is due to the perturbation and \emph{not} because
API sequences were split into two windows due to the added API calls.

As can be seen, the attacks of Uesato et al. \citep{DBLP:conf/icml/UesatoOKO18}
and Ilyas et al. \citep{DBLP:conf/icml/IlyasEAL18} have low effectiveness.
This is due to the fact that those attacks are not suitable for discrete
values of API call types and indices. 

In contrast, we see that our uniform mixing EA score-based attack
has higher attack effectiveness, for a fixed number of queries, even
when used for discrete input (API calls or position indices). This
is due to the fact that the transformations used by EA work with discrete
sequences: mutation (random perturbation) in existing adversarial
candidates and crossover between several candidates. In our EA score-based
attack, we don't use crossover, which might make sense for the NLP
domain (e.g., for compound sentences) but not for API call sequences,
where each program has its own business logic. The self-adaptive search
used by our EA score-based attack also explains why it outperforms
all other score-based attack variants and has better attack effectiveness
than the gradient-based attack used in \citep{DBLP:conf/raid/RosenbergSRE18}
with the same number of queries. Our proposed score-based attack outperforms
existing methods because it maximizes the attack effectiveness for
a fixed number of queries. Note that the number of queries is per
sliding window and not per executable. 

Based on the average malicious sequence length, $avg(length(\boldsymbol{x^{m}}))\approx10,000$,
and the adversarial sliding window size, $k=140$, the average absolute
number of queries per malware executable is \textasciitilde 10,000. 

As expected, the benign perturbation effect on the decision-based
attack effectiveness is the most significant, since without it, the
API types are random. 

While our decision-based attack effectiveness is 10\% lower than the
most effective score-based attacks when using the same budget, it
doesn't require knowledge of the target classifier's confidence score,
making it the only viable attack in some black-box scenarios.

\subsection{\label{subsec:Defenses-and-Mitigation}Defenses and Mitigation Techniques}

To the best of our knowledge, there is currently no published and
evaluated method to make a sequence-based RNN model resistant to adversarial
sequences, beyond a brief mention of adversarial training as a defense
method \citep{DBLP:conf/emnlp/AlzantotSEHSC18,DBLP:journals/corr/abs-1812-05271}.
Adversarial training \citep{Goodfellow14} is the method of adding
adversarial examples, with their non-perturbed label, to the training
set of the classifier. The reason is since adversarial examples are
usually out-of-distribution samples, inserting them into the training
set would cause the classifier to learn the entire training set distribution,
including the adversarial examples.

Adversarial training has several limitations:
\begin{enumerate}
\item It provides a varying level of robustness, depending on the adversarial
examples used.
\item It requires a dataset of adversarial examples to train on. Thus, it
has limited generalization against novel adversarial attacks.
\item It requires retraining the model, incurring significant overhead.
\end{enumerate}
We ran the adversarial attacks, both score-based and decision-based
variants (Section \ref{alg:Logarithmic-Backtracking-Attack}), with
and without benign perturbation (Section \ref{subsec:Benign-Perturbation:-GAN-1})
on the training set, as suggested in \citep{DBLP:conf/iclr/MadryMSTV18}.
For each column in Tables \ref{tab:Attack-Comparison} and \ref{tab:Benign-Perturbation-Attack-1},
we generated 14,000 malicious adversarial examples (50\% generated
by the black-box attack and 50\% by the white-box attack), which replaced
14,000 malicious samples in the original training set. Other sizes
(smaller or larger) resulted in reduced detection rate of the pre-trained
classifier for non-adversarial samples. The adversarial examples were
generated using the same configuration (score/decision-based, random/benign
perturbation, number of queries to generate) as the evaluated attack.
The results were the same across all attack types: The attack effectiveness
remains the same, while the attack overhead and number of queries
were increased by 10-15\%, on average. This is due to the fact that
adversarial training is less effective against random attacks like
ours, because a different stochastic adversarial sequence is generated
every time, making it challenging for the classifier to generalize
from one adversarial sequence to another.

More effective RNN defense methods, including domain specific methods,
e.g., systems that measure CPU usage \citep{DBLP:conf/isi/MoskovitchPGSFPSE07},
contain irregular API call subsequences \citep{Grosse2017a} (such
as the no-op API calls used in this paper), or otherwise assess the
plausibility of our attack \citep{DBLP:conf/sp/PierazziPCC20}, in
order to detect adversarial examples, will be a part of our future
work.

\section{\label{sec:Conclusions-and-Future}Conclusions and Future Work}

In this paper, we presented the first black-box attack (based on the
target classifier's predicted class, with and without its confidence
score, to fit adversary's limited knowledge) that generates \emph{adversarial
sequences while minimizing the number of queries} for the target classifier,
reducing the number of queries by more than 10 times with minimal
loss of attack effectiveness in comparison to the state of the art
attack (\citep{DBLP:conf/raid/RosenbergSRE18}). This query-efficient
approach makes our attack suited to attack cloud models where a large
amount of queries cost money and raise suspicion of an attack, failing
previous attacks. 

We demonstrated those attacks against API call sequence-based malware
classifiers and verified the attack effectiveness for all relevant
common classifiers: RNN variants, feedforward networks, and traditional
machine learning classifiers. These are the first query-efficient
attacks effective against RNN variants and not just CNNs. 

We also evaluated our attacks against four variants of state-of-the-art
score-based query-efficient attacks, modified to fit discrete sequence
input, and showed that our attacks are equal or outperform all of
them. 

Finally, we demonstrated that our attacks are effective even when
multiple feature types, including non-sequential ones, are used (Appendix
\ref{sec:Appendix-D:-Handling}). 

While this paper focuses on API calls and printable strings as features,
the proposed attacks are valid for every modifiable feature type,
sequential or not. Furthermore, our attack is generic and can be applied
to other domains, like text analysis (using word sequences instead
of API calls), as would be demonstrated in our future work.

Our future work will focus on developing domain-specific and domain-agnostic
defense mechanisms against such attacks and analyzing additional self-adaptive
algorithms to find more query-efficient attacks, while evaluating
them on limited knowledge scenarios (e.g., unknown API calls window
size, etc.). 

\newpage{}

\bibliographystyle{plain}
\bibliography{thesis}

\newpage{}

\appendix

\section{\label{sec:Appendix-A:-Tested}Tested Dataset}

We used identical implementation details (e.g., dataset, classifiers'
hyperparameters, etc.) as Rosenberg et al. \citep{DBLP:conf/raid/RosenbergSRE18},
so the attacks can be compared. The details are provided here for
the reader's convenience.

\begin{figure*}[tp]
\begin{centering}
\textsf{\includegraphics[scale=0.66]{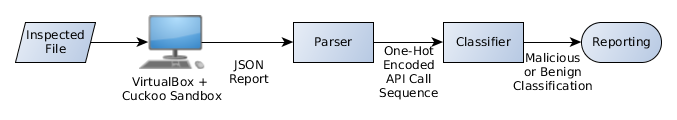}}
\par\end{centering}
\caption{\label{fig:Overview-of-the}Overview of the Malware Classification
Process}
\end{figure*}

An overview of the malware classification process is shown in Figure
\ref{fig:Overview-of-the} (taken from \citep{DBLP:conf/raid/RosenbergSRE18}).

The dataset used is large and includes the latest malware variants,
such as the Cerber and Locky ransomware families. Each malware type
(ransomware, worms, backdoors, droppers, spyware, PUAs, and viruses)
has the same number of samples, to prevent prediction bias towards
the majority class. 20\% of the malware families (such as the NotPetya
ransomware family) were only used in the test set to assess generalization
to an unseen malware family. 80\% of the malware families (such as
the Virut virus family) were distributed between the training and
test sets, to determine the classifier's ability to generalize to
samples from the same family. The temporal difference between the
training set and the test set is six months (i.e., all training set
samples are older than the test set samples, because using a non-time-aware
split may cause a data leak \citep{Pendlebury19}) based on VirusTotal's
'first seen' date. 

The ground truth labels of the dataset were determined by VirusTotal
\citep{VirusTotal}, an online scanning service, which contains more
than 60 different security products. A sample with 15 or more positive
(i.e., malware) classifications from the 60 products is considered
malicious. A sample with zero positive classifications is labeled
as benign. All samples with 1-14 positives were omitted to prevent
false positive contamination of the dataset. Family labels for dataset
balancing were taken from Kaspersky Anti-Virus classifications.

It is crucial to prevent dataset contamination by malware that detects
whether the malware is running in a Cuckoo Sandbox (or on virtual
machines) and if so, quits immediately to prevent reverse engineering
efforts. In those cases, the sample's label is malicious, but its
behavior recorded in Cuckoo Sandbox (its API call sequence) isn't,
due to the malware's anti-forensic capabilities. To prevent such contamination
of the dataset, two countermeasures were used:
\begin{enumerate}
\item Considering only API call sequences with more than 15 API calls, omitting
malware that detects a virtual machine (VM) and quits.
\item Applying Yara rules \citep{YaraRules} to find samples trying to detect
sandbox programs, such as Cuckoo Sandbox, and omitting all such samples. 
\end{enumerate}
One might argue that the evasive malware that applies such anti-VM
techniques is extremely challenging and relevant, however in this
paper we focus on adversarial attacks. Such attacks are generic enough
to work for those evasive types of malware as well, assuming that
other mitigation techniques (e.g., anti-anti-VM), would be applied.
After this filtering and balancing of the benign samples, about 400,000
valid samples remained. The final training set size is 360,000 samples,
36,000 of which serve as the validation set. The test set size is
36,000 samples. All sets are balanced between malicious and benign
samples. Due to hardware limitations, a subset of the dataset was
used (54,000 training samples and test and validation sets of 6,000
samples each). The dataset was representative and maintained the same
distribution as mentioned above.

\section{\label{sec:Appendix-B:-Tested}Tested Malware Classifiers}

\begin{figure}
\begin{centering}
\subfloat[\label{fig:Dynamic-Classifier-Architecture}Dynamic Classifier Architecture]{\includegraphics[scale=0.35]{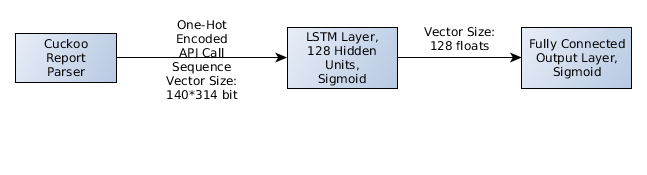}}\hfill{}\subfloat[\label{fig:Hybrid-Classifier-Architecture}Hybrid Classifier Architecture]{\includegraphics[scale=0.35]{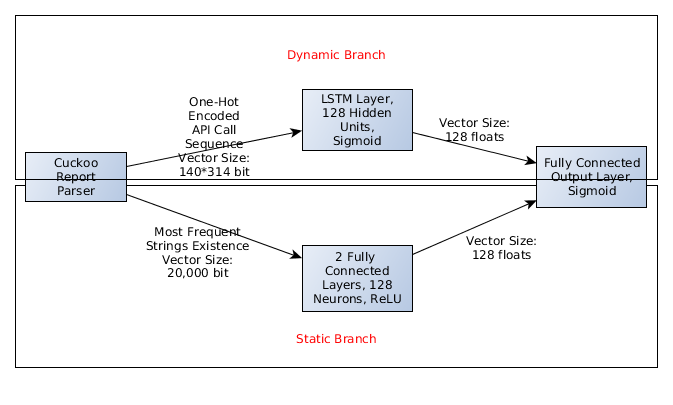}}
\par\end{centering}
\caption{\label{fig:Classifier-Architecture-Overview}Classifier Architecture
Overview}
\end{figure}

As mentioned in Section \ref{subsec:Dataset-and-Target}, we used
the malware classifiers from Rosenberg et al. \citep{DBLP:conf/raid/RosenbergSRE18},
since many classifiers are covered, allowing us to evaluate the attack
effectiveness against many classifier types. The maximum input sequence
length was limited to $k=140$ API calls, since longer sequence lengths,
e.g., $k=1,000$, had no effect on the accuracy, and shorter sequences
were padded with zeros. A zero stands for a null/dummy value API in
our one-hot encoding. Longer sequences are split into windows of $k$
API calls each, and each window is classified in turn. If any window
is malicious, the entire sequence is considered malicious. Thus, the
input of all of the classifiers is a vector of $k=140$ API call types
in one-hot encoding, using 314 bits, since there were 314 monitored
API call types in the Cuckoo reports for the dataset. The output is
a binary classification: malicious or benign. An overview of the LSTM
architecture is shown in Figure \ref{fig:Dynamic-Classifier-Architecture}. 

The Keras \citep{Keras} implementation was used for all neural network
classifiers, with TensorFlow used for the backend. XGBoost \citep{XGBoost}
and scikit-learn \citep{SciKitLearn} were used for all other classifiers. 

The loss function used for training was binary cross-entropy. The
Adam optimizer was used for all of the neural networks. The output
layer was fully connected with sigmoid activation for all neural networks.
For neural networks, a rectified linear unit, $ReLU(x)=max(0,x)$,
was chosen as an activation function for the input and hidden layers
due to its rapid convergence compared to $sigmoid()$ or $\tanh()$,
and dropout was used to improve the generalization potential of the
network. A batch size of 32 samples was used. The classifiers also
have the following classifier specific hyperparameters:
\begin{itemize}
\item DNN - two fully connected hidden layers of 128 neurons, each with
ReLU activation and a dropout rate of 0.2.
\item CNN - 1D ConvNet with 128 output filters, a stride length of one,
a 1D convolution window size of three, and ReLU activation, followed
by a global max pooling 1D layer and a fully connected layer of 128
neurons with ReLU activation and a dropout rate of 0.2.
\item RNN, LSTM, GRU, BRNN, BLSTM, and bidirectional GRU - a hidden layer
of 128 units, with a dropout rate of 0.2 for inputs and recurrent
states.
\item Deep LSTM and BLSTM - two hidden layers of 128 units, with a dropout
rate of 0.2 for inputs and recurrent states in both layers.
\item Linear SVM and logistic regression classifiers - a regularization
parameter of C=1.0 and an L2 norm penalty.
\item Random forest classifier - 10 decision trees with unlimited maximum
depth and the Gini criteria for choosing the best split.
\item Gradient boosted decision tree - up to 100 decision trees with a maximum
depth of 10 each.
\end{itemize}
The classifiers' performance was measured using the accuracy ratio,
which gives equal importance to both false positives and false negatives
(unlike precision or recall). The false positive rate of the classifiers
varied between 0.5-1\%. The false positive rate was chosen to be on
the high end of production systems. A lower false positive rate would
mean lower recall either, due to the trade-off between false positive
rate and recall, thereby making our attacks even more effective. 

The performance of the classifiers is shown in Table \ref{tab:Classifier-Performance}.
The accuracy was measured on the test set, which contains 36,000 samples.

\begin{table}[htbp]
\caption{\label{tab:Classifier-Performance}Classifier Performance}

\centering{}%
\begin{tabular}{|c|c|}
\hline 
Classifier Type & Accuracy (\%)\tabularnewline
\hline 
\hline 
LSTM & 98.26\tabularnewline
\hline 
Deep LSTM & 97.90\tabularnewline
\hline 
GRU & 97.32\tabularnewline
\hline 
Bidirectional GRU & 98.04\tabularnewline
\hline 
1D CNN & 96.42\tabularnewline
\hline 
Random Forest & 91.90\tabularnewline
\hline 
SVM & 86.18\tabularnewline
\hline 
Logistic Regression & 89.22\tabularnewline
\hline 
Gradient Boosted Decision Tree & 91.10\tabularnewline
\hline 
\end{tabular}
\end{table}

As can be seen in Table \ref{tab:Classifier-Performance}, the LSTM
variants are the best malware classifiers, in terms of accuracy, and,
as shown in Section \ref{subsec:Attack-Performance}, BLSTM is also
one of the classifiers most resistant to the proposed attack.

\section{\label{sec:Appendix-C:-Benign}Benign Perturbation GAN Comparison}

To implement the benign perturbation GAN, we tested several GAN types,
using Texygen \citep{DBLP:conf/sigir/ZhuLZGZWY18} with its default
parameters. We use maximum likelihood estimation (MLE) training as
the pretraining process for all baseline models except GSGAN, which
requires no pretraining. In pretraining, we first train 80 epochs
for a generator, and then train 80 epochs for a discriminator. The
adversarial training comes next. In each adversarial epoch, we update
the generator once and then update the discriminator for 15 mini-batch
gradients. Due to memory limitations, we generated only one sliding
window of 140 API calls, each with 314 possible API call types, in
each iteration (that is, generating\textbf{ $\boldsymbol{w^{b}}$}
and not \textbf{$\boldsymbol{x^{b}}$} as described in Algorithm \ref{alg:Logarithmic-Backtracking-Attack}). 

We tested several GAN implementations with discrete sequence output.
We trained our GAN using a benign hold-out set (3,000 sequences).
Next, we run Algorithm \ref{alg:Logarithmic-Backtracking-Attack}
(logarithmic backtracking transformation with benign perturbation)
on the 3,000 API call traces generated by the GAN. Finally, we used
the benign test set (3,000 sequences) as the GAN's test set. The results
for the LSTM classifier are shown in Table \ref{tab:Benign-Perturbation-Attack-2}
(the results for other classifiers, which are not shown due to space
limits, are similar).

\begin{table*}
\caption{\label{tab:Benign-Perturbation-Attack-2}Benign Perturbation Attack
Performance}

\centering{}%
\begin{tabular}{|>{\centering}p{0.25\textwidth}|>{\centering}p{0.2\textwidth}|>{\centering}p{0.2\textwidth}||>{\centering}p{0.15\textwidth}|}
\hline 
GAN Type & Attack Effectiveness {[}\%{]} & Added API Calls {[}\%{]} & Queries Used\tabularnewline
\hline 
\hline 
None (Random Perturbation) & 21.25 & 27.94 & 119.40\tabularnewline
\hline 
\hline 
SeqGAN \citep{DBLP:conf/aaai/YuZWY17} & \textbf{89.39} & \textbf{12.82} & \textbf{17.73}\tabularnewline
\hline 
\hline 
TextGAN \citep{DBLP:conf/icml/ZhangGFCHSC17} & 74.53 & 16.74 & 30.58\tabularnewline
\hline 
\hline 
GSGAN \citep{DBLP:journals/corr/KusnerH16} & 88.19 & 14.06 & 20.43\tabularnewline
\hline 
\hline 
MaliGAN \citep{DBLP:journals/corr/CheLZHLSB17} & 86.67 & 15.12 & 22.74\tabularnewline
\hline 
\end{tabular}
\end{table*}

We can see from the results presented in the table that SeqGAN outperforms
all of the other models in all of the measured factors, due to its
RL-based ability to pass gradient updates between the generator and
discriminator parts of the GAN. We also see that, as expected, a random
perturbation is less effective than a benign perturbation, regardless
of the type of GAN used.

\section{\label{sec:Appendix-D:-Handling}Handling Multiple Feature Types
and Hybrid Classifiers}

Combining several types of features can make the classifier more resistant
to adversarial examples against a specific feature type. For instance,
some real-world next generation anti-malware products are hybrid classifiers,
combining both static and dynamic features for a better detection
rate. An extension of our attack, enabling it to handle hybrid classifiers,
is straightforward: attacking \emph{each feature type in turn} using
Algorithm \ref{alg:Logarithmic-Backtracking-Attack}. If the attack
against a feature type fails, we continue and attack the next feature
type with the modified binary until a benign classification by the
target model is achieved or all feature types have been (unsuccessfully)
attacked. We used the same hybrid malware classifier specified in
Appendix \ref{sec:Appendix-C:-Benign}, for which the input consists
of both an API call sequence and the most frequent 20,000 printable
strings inside the PE file as Boolean features (exist or not). 

While there are more complex static features (e.g., \citep{DBLP:journals/corr/abs-1804-04637}),
we chose printable strings, easy to modify features that have been
used by many classifiers \citep{DBLP:journals/jmlr/KolterM06}, as
a concrete example of the multi-feature use case, to show that the
suggested attack works not only against RNNs, but also against other
classifiers, making it more generic.

We evaluated the performance of our decision-based, linear iteration,
benign perturbation attack. When attacking only the API call sequences
using the hybrid classifier, without modifying the static features
of the sample, the attack effectiveness decreases to 23.76\%. This
is much lower than the attack effectiveness of 89.67\% obtained for
a classifier trained only on the dynamic features, meaning that the
attack was mitigated by the use of additional static features. When
attacking only the printable string features (again, assuming that
the attacker has the knowledge of $D'=D$, which contains the printable
strings being used as features by the hybrid classifier), the attack
effectiveness is 28.25\%. This is much lower than the attack effectiveness
of 88.31\% obtained for a classifier trained only on the static features.
Finally, the multi-feature attack's effectiveness for the hybrid model
was 90.06\%. Other types of classifiers and attacks provided similar
results. They are not presented here due to space limits.

To summarize, we have shown that while the use of hybrid models decreases
the specialized attacks' effectiveness, our suggested hybrid attack
performs well, with high attack effectiveness. While not shown due
to space limits, the attack overhead isn't significantly affected.

\section{\label{sec:Appendix-E:-Tested}Tested Hybrid Malware Classifiers}

As mentioned in Appendix \ref{sec:Appendix-D:-Handling}, we used
the hybrid malware classifier used in \citep{DBLP:conf/raid/RosenbergSRE18},
with printable strings inside a PE file as our static features. Strings
can be used, e.g., to statically identify loaded DLLs and called functions,
and recognize modified file paths and registry keys, etc. Our architecture
for the hybrid classifier, shown in Figure \ref{fig:Hybrid-Classifier-Architecture},
is:
\begin{enumerate}
\item A static branch that contains an input vector of 20,000 Boolean values:
for each of the 20,000 most frequent strings in the entire dataset,
do they appear in the file or not? This is analogous to a similar
procedure used in NLP which filters the least frequent words in a
language.
\item A dynamic branch that contains an input vector of 140 API calls (each
of which is one-hot encoded) inserted into an LSTM layer of 128 units
and a sigmoid activation function, with a dropout rate of 0.2 for
inputs and recurrent states. This vector is inserted into two fully
connected layers with 128 neurons, a ReLU activation function, and
a dropout rate of 0.2 each. 
\end{enumerate}
The 256 outputs of both branches are inserted into a fully connected
output layer with a sigmoid activation function. Therefore, the input
of the classifier is a vector containing 20,000 Boolean values and
140 one-hot encoded API call types, and the output is malicious or
benign classification. The training set size was reduced by half in
comparison to the training set specified in Appendix \ref{sec:Appendix-A:-Tested}
(while keeping the same dataset structure) due to the larger memory
requirements for training a classifier with more features. All other
hyperparameters are the same as those mentioned in Appendix \ref{sec:Appendix-B:-Tested}.

A classifier using only the dynamic branch (Figure \ref{fig:Dynamic-Classifier-Architecture})
achieves 92.48\% accuracy on the test set (this is different from
the results presented in Table \ref{tab:Classifier-Performance},
due to the smaller training set), a classifier using only the static
branch attains 96.19\% accuracy, and a hybrid model that uses both
branches (Figure \ref{fig:Hybrid-Classifier-Architecture}) obtains
96.94\% accuracy, meaning that using multiple feature types improves
the accuracy.
\end{document}